\documentclass[pre,twocolumn,amsmath]{revtex4}
\usepackage{graphicx}
\begin{document}
\title{Logarithmic roughening in a growth process with edge evaporation}

\author{Haye Hinrichsen}
\affiliation{Theoretische Physik, Fachbereich 8, Universit{\"a}t
	     Wuppertal, 42097 Wuppertal, Germany}

\begin{abstract}
Roughening transitions are often characterized by unusual scaling properties. As an example we investigate the roughening transition in a solid-on-solid growth process with edge evaporation [Phys. Rev. Lett. 76, 2746 (1996)], where the interface is known to roughen logarithmically with time. Performing high-precision simulations we find appropriate scaling forms for various quantities. Moreover we present a simple approximation explaining why the interface roughens logarithmically.
\end{abstract}


\maketitle
\parskip 2mm 
\def\growth{\gamma} 

\section{Introduction}

A large variety of models for surface growth is known to
display universal scaling laws~\cite{Meakin,KrugSpohn,Dietrich,Barabasi,Krug}. In most cases one observes simple power-law scaling, i.e., the width~$W$ of an initially flat interface grows with time as
\begin{equation}
\label{Roughening}
W \sim t^\growth \,,
\end{equation}
where $\growth$ is the so-called growth exponent. In a finite system the width eventually saturates at a finite value $W_{\rm sat}$, which grows with the linear system size $L$ as
\begin{equation}
\label{Saturation}
W_{\rm sat} \sim L^\alpha \,,
\end{equation}
where $\alpha$ is the roughness exponent. Both asymptotic power laws can be combined in a single scaling form
\begin{equation}
\label{StandardScaling}
W(L,t) = L^\alpha \, f(t/L^z)\,,
\end{equation}
where $z=\alpha/\gamma$ is the dynamic exponent and $f$ is a scaling function with an appropriate asymptotic behavior. This type of power-law scaling, also known as Family-Vicsek scaling~\cite{FamilyVicsek}, is generic for a large number of self-similar growth processes. However, in some cases the interface was found to roughen {\em logarithmically} with time. This may happen, for example, when a system undergoes a roughening transition. The purpose of this paper is to study a simple model with such a logarithmic scaling behavior in more detail. To this end we consider a class of solid-on-solid growth models introduced a few years ago by Alon {\it et al.}~\cite{Alon,Alon2} which have the special feature that atoms can only evaporate at the edges of terraces. The models exhibit a robust phase transition from a non-moving smooth phase to a moving phase, where the interface roughens continuously. Remarkably, this roughening transition takes place even in one spatial dimension. At the critical point the interface width was found to increase logarithmically with time. However, so far it was impossible to unfold the complete scaling picture since previous numerical simulations were not accurate enough~\cite{Alon2}. Here we close this gap by performing high-precision simulations based on a parallelized code. As a main result we find that Eq.~(\ref{StandardScaling}) has to be replaced by a logarithmic scaling law of the form
\begin{equation}
\label{LogScaling}
W^2(L,t) = \lambda \ln L + f(t/L^z)\,,
\end{equation}
Moreover, we present a simple approximation which explains why the interface roughens logarithmically.

\section{Growth model with evaporation at the edges of plateaus}     
\label{RougheningSection}                    

\begin{figure}
\includegraphics[width=40mm]{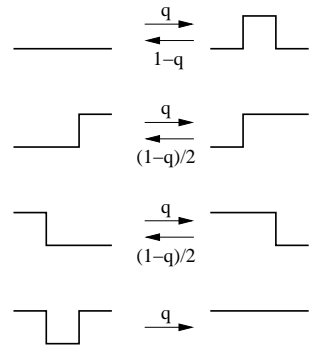}
\caption{
\label{FIGRULES}
Dynamic rules for deposition $(\rightarrow)$ and
evaporation $(\leftarrow)$ of the restricted model 
introduced in Ref.~\cite{Alon}. Note that evaporation from 
the middle of plateaus is not allowed.
}
\end{figure}

The growth models investigated in Ref.~\cite{Alon} are defined as solid-on-solid deposition-evaporation processes with the special property that atoms can only evaporate at the edges of terraces. The models are defined on a $d$-dimensional lattice with $L$ sites and periodic boundary conditions, where each site $i$ is associated with an integer variable~$h_i$. In Ref.~\cite{Alon} two variants of models have been considered. Here we will focus on the physically motivated {\em restricted} model in one dimension, where the heights at neighboring sites obey the inequality
\begin{equation}
\label{RSOS}
|h_i-h_{i+1}| \leq 1\,.
\end{equation}
The model evolves by random-sequential updates, i.e., a site $i$ is randomly selected and one of the following moves is carried out (see Fig.~\ref{FIGRULES}). With probability $q$ an atom is deposited at site $i$, increasing $h_i$ by $1$. Otherwise one of the nearest neighbors is randomly selected. If the selected neighbor is at a lower height, indicating the edge of a terrace, an atom is removed from site $i$. In both cases a move is abandoned if the resulting configuration would violate the condition~(\ref{RSOS}). Each attempted update corresponds to an average time increment $\Delta t=1/L$.

In the limit $L\to \infty$ the model has the following phenomenological properties. If the growth rate $q$ is small, the interface is flat and pinned to a spontaneously selected bottom layer. Small islands will occasionally grow on top of this bottom layer but will quickly be eliminated by desorption at the islands edges so that the system eventually reaches a stationary state characterized by a finite width. As $q$ is increased, more and more islands on top of the bottom layer are produced, until above $q_c$, the critical value of $q$, the islands merge and the interface starts moving. Approaching $q_c$ from below, the stationary width is found to diverge logarithmically.

\begin{figure}
\includegraphics[width=88mm]{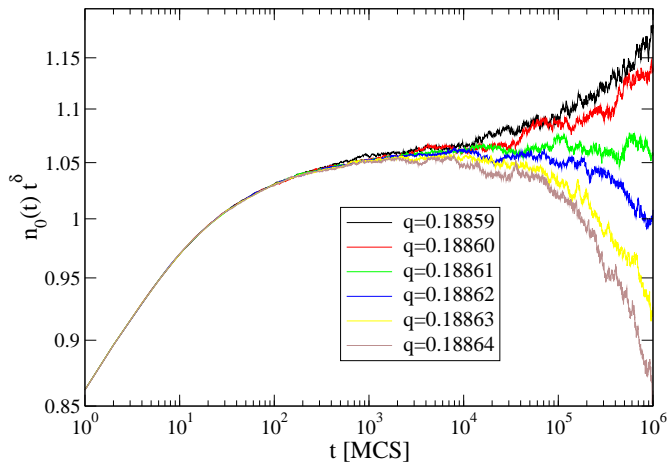}
\caption{
\label{FIGBASE}
Density of exposed sites at the bottom layer
$n_0(t)$ times $t^\delta$ in a system with 32768 sites
averaged over up to 1000 independent runs for different values of q.
}
\end{figure}
In Refs.~\cite{Alon,Alon2} it was shown that the roughening transition in this class of models is driven by a directed percolation (DP) transition~\cite{Hinrichsen00a} at the spontaneously selected bottom layer. This means that the density of exposed sites at the bottom layer $n_0$ can be interpreted as the density of active sites in a DP process. Therefore, at criticality this density is expected to decay as
\begin{equation}
\label{CPDecay}
n_0 \sim t^{-\delta} \,,
\end{equation}
where $\delta =\beta/\nu_\perp \simeq 0.1595$ is one of the dynamic critical exponents of DP in 1+1 dimensions. Similarly the critical behavior at the first few layers above the bottom layer can be described in terms of unidirectionally coupled DP processes~\cite{Coupled}. Roughly speaking each layer can be associated  with a DP process which is coupled to the layer below. Effective couplings in opposite direction, which certainly exist in the model defined above, turn out to be irrelevant. Field-theoretic renormalization group studies revealed that the scaling exponents $\nu_\parallel$ and $\nu_\perp$ (and hence their ratio $z=\nu_\parallel/\nu_\perp$) are the same at all levels, while the exponents associated with the order parameters decrease with increasing height above the bottom layer. Thus the concept of unidirectionally coupled DP successfully explains the critical behavior at the first few layers above the bottom layer. However, it cannot predict the critical properties of the interface as a whole, especially the scaling of the interface width. 

\section{Numerical simulations}     
\label{NumericalSimulations}                    

\subsection{Scaling properties at the critical point}
\label{decay}

%
%
%
\begin{figure}
\includegraphics[width=82mm]{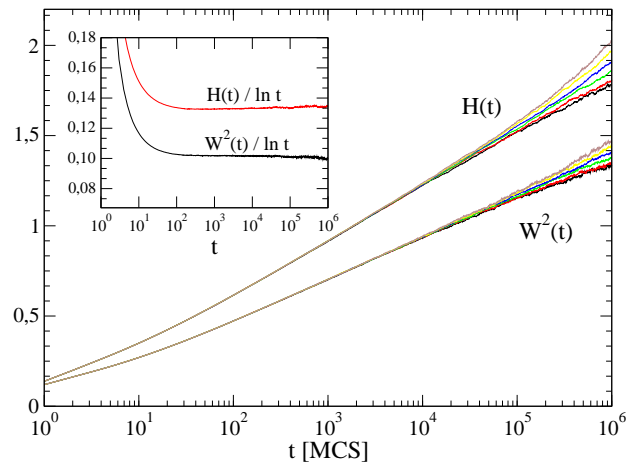}
\caption{
\label{FIGWIDTH}
The squared interface width $W^2$ and the average height $H$ as functions of time in a log-lin plot. The simulation parameters are the same as in Fig.~\ref{FIGBASE}. The inset shows that both quantities increase logarithmically at the critical point. Small deviations for $t>10^5$ can be traced back to the onset of finite size effects. 
}
\end{figure}
Performing high-precision simulations on a parallel computer we first used Eq.~(\ref{CPDecay}) to estimate the critical threshold. As shown in Fig.~\ref{FIGBASE} we obtain the result
\begin{equation}
\label{CriticalPoint}
q_c = 1.8861(1)
\end{equation}
which is much more accurate than the previous estimate reported in Ref.~\cite{Alon}. Moreover, our simulations confirm that the dynamics at the bottom layer is indeed driven by a DP process.

At the critical point $q=q_c$ the average height
\begin{equation}
H=\frac{1}{L}\sum_{i=1}^Lh_i
\end{equation}
and the squared interface width 
\begin{equation}
W^2=\frac1L\sum_{i=1}^L(h_i-H)^2
\end{equation}
are found to increase {\em logarithmically} with time as
\begin{eqnarray}
W^2(t) &\simeq& \tau \ln t\,,\\
H(t) &\simeq& \sigma \ln t\,,
\end{eqnarray}
where $\tau=0.102(3)$ and $\sigma=0.133(3)$ (see Fig.~\ref{FIGWIDTH}). Thus, apart from different prefactors, the squared width and the mean height show the same type of logarithmic behavior. We note that in a previous work the width was erroneously conjectured to increase as $W(t)\sim (\ln t)^{0.43}$~\cite{Alon2}.

\subsection{Finite-size simulations}
\label{fs}

%
%
\begin{figure}
\includegraphics[width=56mm]{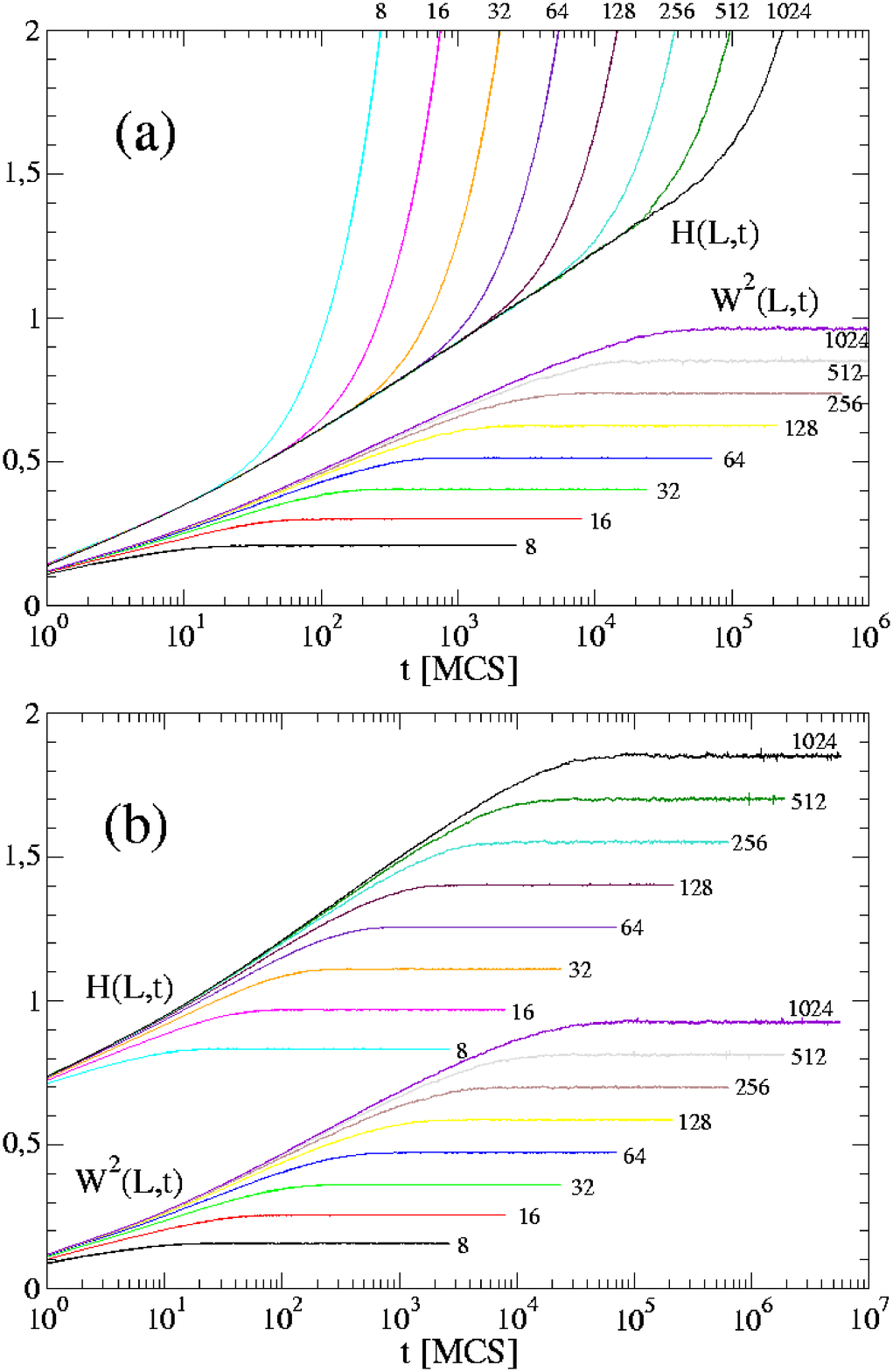} 
\caption{
\label{FIGFSRAW}
Finite size simulations:
Squared width $W^2(L,t)$ and mean height $H(L,t)$ averaged over at least 500 independent runs in a finite system with $L=8,16,32,\ldots,1024$ sites  for (a) periodic and (b) fixed boundary conditions. In the lower panel the curves for $H(t)$ have been shifted vertically.
}
\end{figure}
In a finite system the critical behavior depends on the type of the boundary condition. In what follows we consider (a) periodic and (b) Dirichlet boundary conditions, where the sites at the boundary are fixed at zero height. As shown in Fig.~\ref{FIGFSRAW}, the average height $H(L,t)$  and the squared width $W^2(L,t)$ first increase logarithmically until finite-size effects become relevant and the system crosses over to a different regime, where the width saturates. Measuring the saturation levels we find that 
\begin{equation}
W^2_{\rm sat}(L) \simeq A_{p,f} + \lambda_{p,f} \ln L\,,
\end{equation}
where the indices $p,f$ stand for periodic and fixed boundary conditions, respectively. Our best estimates are:
\begin{equation}
\begin{split}
A_p&=-0.16(1)\,, \qquad \lambda_p=0.161(2)\\
A_f&=-0.21(1)\,, \qquad \lambda_f=0.163(2)
\end{split}
\end{equation}
Apparently the amplitude $\lambda$ is universal in the sense that it does not depend on the choice of the boundary condition. 

Unlike the width, the interface height does only saturate if fixed boundary conditions are used.
Here we find a similar formula
\begin{equation}
\label{Hfixed}
H_{\rm sat}(L) \simeq B_f + \mu_f \ln L\,,
\end{equation}
where
\begin{equation}
B_f=-0.24(1)\,, \qquad \mu_f = 0.212(10)\,.
\end{equation}
\begin{figure}
\includegraphics[width=90mm]{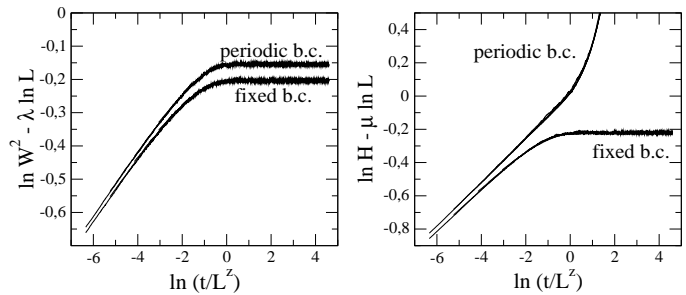}
\caption{
\label{FIGFSCOL}
Finite size simulations: Data collapse of the curves shown in Fig.~\ref{FIGFSRAW} according to the scaling forms~(\ref{FSScalingForms}) for $L=32,64,\ldots,1024$. In order to eliminate deviations due to initial transients all data points for $t<100$ have been discarded.
}
\end{figure}

In order to collapse the curves shown in Fig.~\ref{FIGFSRAW}, recall that the roughening transition in this model is driven by a 1+1-dimensional DP process at the bottom layer, which is characterized by a dynamic exponent
\begin{equation}
z=\nu_\parallel/\nu_\perp\simeq 1.5807\,.
\end{equation}
Moreover it has been shown that the critical dynamics at the first few layers is the same as in a sequence of unidirectionally coupled DP processes, which are all characterized by the {\em same} dynamic exponent $z$. Therefore, it is reasonable to assume that the roughening transition as a whole is characterized by the dynamic exponent of DP so that finite-size scaling functions can only depend on the scaling-invariant ratio $t/L^z$. Thus, the expected scaling forms read
\begin{equation}
\label{FSScalingForms}
\begin{split}
W^2(L,t) \simeq \lambda \ln L + f(t/L^z) \,,\\
H(L,t)   \simeq \mu     \ln L + g(t/L^z) \,,
\end{split}
\end{equation}
where $f(\xi)$ and $g(\xi)$ are scaling functions with the asymptotic behavior
\begin{equation}
f(\xi) = \frac{\lambda}{z}\ln \xi \,, \qquad
g(\xi) = \frac{\mu}{z}\ln \xi 
\end{equation}
for $\xi \to 0$ and $f(\xi)=A$ and $g(\xi)=B$ for $\xi \to \infty$.
In order to verify these scaling forms we plot $W^2(L,t)-\lambda \ln L$ and $H(L,t)-\mu \ln L$ versus $t/L^z$. As shown in Fig.~\ref{FIGFSCOL}, we obtain excellent data collapses. 

\subsection{Off-critical simulations}
\label{off}
%
%
%
\begin{figure}
\includegraphics[width=90mm]{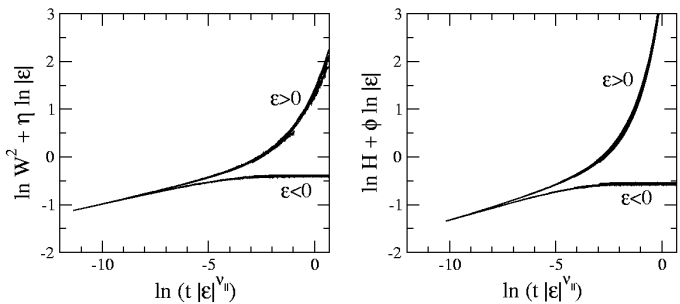} 
\caption{
\label{FIGOFF}
Off-critical simulations for $\epsilon=\pm 0.0001$, $\pm0.0002$, $\ldots,\pm 0.0128$ on a chain with 32768 sites averaged over 200 independent runs. The figure shows data collapses according to the scaling forms~(\ref{OffScalingForms}). In order to eliminate deviations due to initial transients all data points for $t<100$ have been removed.
}
\end{figure}

A similar situation is encountered in off-critical simulations. In the subcritical regime $q<q_c$ the width and the average height saturate at
\begin{equation}
\label{OffScalingForms}
\begin{split}
W^2_{\rm sat}(\epsilon) &= C - \eta \ln |\epsilon|\,, \\
H_{\rm sat}(\epsilon) &= D - \phi \ln |\epsilon|\,, \\
\end{split}
\end{equation}
where $\epsilon=q-q_c$ denotes the distance from criticality. Following the same arguments as in the previous subsection, the conjectured scaling forms read
\begin{equation}
\label{OffScalingForm}
\begin{split}
W^2(\epsilon,t) &= - \eta \ln |\epsilon| + F(t |\epsilon|^{\nu_\parallel})\,,\\
H(\epsilon,t) &= - \phi \ln |\epsilon| + G(t |\epsilon|^{\nu_\parallel})\,,
\end{split}
\end{equation}
where $\nu_\parallel \simeq 1.7338$ is the temporal scaling exponent of DP in 1+1 dimensions. These scaling forms are expected to hold not only below but also above criticality. In fact, plotting $W^2(\epsilon,t) +\eta \ln |\epsilon|$ and $H(\epsilon,t) +\phi \ln |\epsilon|$versus $t |\epsilon|^{\nu_\parallel}$ (see Fig.~\ref{FIGOFF}), the best data collapses are obtained for 
\begin{equation}
\eta=0.173(10) \,,\qquad \phi=0.230(10)\,,
\end{equation}
while the constants are estimated by
\begin{equation}
C=-0.40(1) \,,\qquad D=-0.54(1)\,.
\end{equation}
%
%
\subsection{Relations among the amplitudes}
\label{Amplitudes}
%
%
Obviously the scaling forms (\ref{FSScalingForms}) and (\ref{OffScalingForms}) suggest that the {\em exponentiated} quantities $\exp(W^2)$ and $\exp(H)$ display ordinary power-law scaling. This would imply that the results for finite-size and off-critical scaling can be combined in a single scaling form, e.g.,
\begin{equation}
\begin{split}
\exp\bigl[W^2(\epsilon,L,t)\bigr] &= L^\lambda \, \tilde{F}(\epsilon L^{1/\nu_\perp},t/L^z)\,,\\
\exp\bigl[H(\epsilon,L,t)\bigr] &= L^\mu \, \tilde{G}(\epsilon L^{1/\nu_\perp},t/L^z)\,.
\end{split}
\end{equation}
This in turn would imply that the six amplitudes $\lambda,\mu,\tau,\sigma,\eta,\phi$ introduced above, which now play the role of critical exponents, are no longer independent. Only two of them, say $\eta$ and $\phi$, are independent while the others can be expressed as
\begin{equation}
\begin{split}
\lambda &= \eta/\nu_\perp \,,\\
\mu &= \phi/\nu_\perp \,,\\
\tau &= \eta/\nu_\parallel \,,\\
\sigma &= \phi/\nu_\parallel \,,
\end{split}
\end{equation}
where $\nu_\perp \simeq 1.0968$ and $\nu_\parallel \simeq 1.7338$.
As shown in Table~\ref{Table} these relations are compatible with the numerical results.

\begin{table}
\begin{tabular}{|c|c|c|}\hline
\ \ \ quantity \ \ \ & 	\ \ \ 	amplitude \ \ \ & \ \ \ estimate\ \ \  \\
\hline\hline
$W^2(t)$ &		$\tau$ &		$0.102(3)$ \\
$H(t)$   &		$\sigma$ &		$0.133(3)$\\
\hline
$W^2(L,t)$	&	$\lambda_p$ &		$0.161(2)$ \\
$W^2(L,t)$    &		$\lambda_f$ &		$0.163(2)$ 	\\
$H(L,t)  $	&	$\mu_p$ &		$0.208(10)$ \\
$H(L,t)  $ &		$\mu_f$	&		$0.212(10)$\\
\hline
$W^2(\epsilon,t)$ &	$\eta$	&		$0.173(10)$  \\
$H(\epsilon,t) $ &	$\phi$	&		$0.230(10)$  \\
\hline
\end{tabular}
\caption{\label{Table} Summary of the estimated amplitudes.}
\end{table}
%
%

\section{Simple approximation}                         
\label{Approximation}                    

Why does the interface  roughen logarithmically at the transition? To answer this question we present a simple approximation for the average interface height $H(L)$ in a finite 1+1-dimensional system at criticality. We consider fixed boundary conditions, i.e., the interface is pinned to the bottom layer at the two boundary sites. In between the interface may return to the bottom layer several times, dividing the chain into intervals of different lengths $\ell_1,\ell_2,\ldots$, as sketched in Fig.~\ref{FIGAPPROX}. The approximation relies on the following assumptions:
\begin{figure}
\includegraphics[width=85mm]{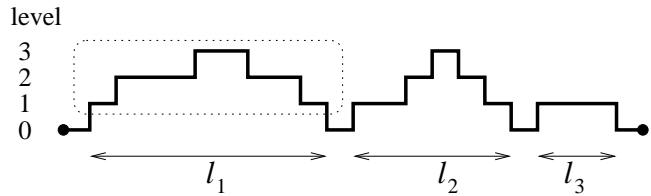}
\caption{
\label{FIGAPPROX}
Approximation of the average height by considering islands of size $\ell_1,\ell_2,\ell_3,\ldots$ (see text).
}
\end{figure}
\begin{enumerate}
\item The probability distribution for the interval sizes~$\ell$ is the same as the size distribution of inactive islands in a DP process, which is known to decrease as $P(\ell) \sim \ell^{\beta/\nu_\perp-2}\simeq\ell^{-0.25}$.
\item As we are using fixed boundary conditions, we will assume that this probability distribution is cut off at the system size, i.e., $P(\ell)=0$ for $\ell \geq L$.
\item The sites, where the interface touches the bottom layer, can be regarded as fixed boundary conditions for the dynamics taking place at the following layer.
\end{enumerate}
While the first two assumptions can be justified by numerical checks, the third assumption is crucial since the sites at the bottom layer evolve in time and can only be regarded as fixed in an approximate sense. However, using this approximation we can derive a recursion relation since each island (for example the island encircled by the dotted line in Fig.~\ref{FIGAPPROX}) can be regarded as an independent growth process taking place in a finite system of size~$\ell$ with fixed boundary conditions at the following layer. Obviously, the average height of such an island will be $1+H(\ell)$. In order to calculate the average height of the entire interface $H(L)$, one has to add up all these contributions weighted by the island size $\ell$. Replacing sums by integrals, this leads to the integral equation
\begin{equation}
H(L) = \frac{\int_0^L d\ell \, \ell P(\ell) \bigl[1+H(\ell)\bigr]}
{\int_0^L d\ell \, \ell P(\ell)}\,.
\end{equation}
Shifting the denominator to the l.h.s. and differentiating with respect to $L$ we obtain
\begin{equation}
\label{DGL2}
\frac{dH(L)}{dL}\, \int_0^L d\ell \, \ell P(\ell) = L P(L)\,,
\end{equation}
or equivalently
\begin{equation}
\frac{dH(L)}{dL} = \frac{d}{dL} \,\ln\,\int_0^L d\ell \, \ell P(\ell)\,,
\end{equation}
leading to the solution
\begin{equation}
\begin{split}
H(L) &= const + \ln \int_0^L d\ell \, \ell P(\ell)\\
     &= const + \frac{\beta}{\nu_\perp} \ln L \,.
\end{split}
\end{equation}
Thus the approximation correctly reproduces the logarithmic increase of the average height in a finite critical system. However, the predicted amplitude $\mu=\beta/\nu_\perp \simeq 0.25$ differs from the measured value $\mu \approx 0.21$. 

A similar argument can be used to explain the logarithmic increase of the squared width. To this end we first calculate the second moment of the height %
\begin{equation}
M_2(L)=\frac1L\sum_{i=1}^Lh_i^2\,.
\end{equation}
Using the same assumptions as before, each island of size~$\ell$ contributes to $M_2(L)$ with a term 
\begin{equation}
M_2(\ell) \Bigr|_{h \to h+1}
 = 1+2H(\ell)+M_2(\ell)\,.
\end{equation}
Replacing sums by integrals we obtain the integral equation
\begin{equation}
M_2(L) = \frac{\int_0^L d\ell \, \ell P(\ell) \bigl[1+2H(\ell)+M_2(\ell)\bigr]}
{\int_0^L d\ell \, \ell P(\ell)}\,,
\end{equation}
leading to the differential equation
\begin{equation}
\label{DGL3}
\frac{dM_2(L)}{dL}\, \int_0^L d\ell \, \ell P(\ell) = L P(L) + 2LP(L)H(L)\,.
\end{equation}
In order to remove the mixed term on the r.h.s. we differentiate the squared width $W^2(L)=M_2(L)-H^2(L)$ with respect to $L$ and combine the result with Eq.~(\ref{DGL2}):
\begin{equation}
\frac{d[W^2(L)-M_2(L)]}{dL} \, \int_0^L d\ell \, \ell P(\ell) = - 2LP(L)H(L)\,.
\end{equation}
Inserting Eq.~(\ref{DGL3}) the mixed term drops out and we obtain the differential equation
\begin{equation}
\frac{dW^2(L)}{dL}\, \int_0^L d\ell \, \ell P(\ell) = L P(L)\,,
\end{equation}
which has exactly the same form as Eq.~(\ref{DGL2}). Therefore, $W^2(L)$ and $H(L)$ will only differ by a constant. Again the approximation correctly reproduces the logarithmic increase but cannot predict the precise value of the amplitude.

\section{Conclusions}                         
\label{Conclusions}                    

In the present paper we have analyzed the scaling behavior at the roughening transition of a restricted solid-on-solid growth process with evaporation at the edges of terraces. Restricting to the case of a one-dimensional substrate we have carried out extensive Monte Carlo simulations, exceeding the temporal range of previous numerical studies by three decades. The numerical results confirm that the transition is indeed driven by a directed percolation process at the bottom layer. Furthermore, they clearly show that squared interface width $W^2(t)$ and the average height above the bottom layer $H(t)$ increase logarithmically with time. Analyzing  numerical data from finite-size and off-critical simulations we have postulated appropriate scaling forms which generate excellent data collapses. These scaling forms can be interpreted in such a way that the exponentiated quantities $\exp(W^2)$ and $\exp(H)$ obey ordinary power-law scaling, which allows one to derive various relations among the amplitudes. Moreover, we have presented a simple approximation which explains why $W^2$ and $H$ grow logarithmically.

In contrast to the expectations expressed in~\cite{Alon2}, we cannot find evidence for several competing length scales. Moreover, there is no evidence for logarithmic scaling based on local scale invariance~\cite{Tang}, rather the system shows ordinary scaling after an appropriate redefinition (exponentiation) of the order parameters.

Our results suggest that any roughening transition, which is driven by a phase transition from an active into an absorbing state at the bottom layer, should exhibit this type of logarithmic roughening. An important example is a recently introduced model for dimer adsorption and desorption~\cite{Dimer}, where the transition is driven by a parity-conserving transition at the bottom layer.

\noindent {\bf Acknowledgements:} The simulations were partly performed on the ALiCE parallel computer at the IAI in Wuppertal. I would like to thank B. Orth and G. Arnold for technical support.


\end{document}